\documentstyle[multicol,pre,aps]{revtex}

\begin{document}

\draft
\title{ Generalizing the Debye-H\"uckel equation\\in terms of density functional integral }
\author{H. Frusawa\thanks{Electronic address: {\ttfamily
furu@exp.t.u-tokyo.ac.jp}} and R. Hayakawa}
\address{Department of Applied Physics, University of Tokyo, Bunkyo-ku, Tokyo 113-8656, Japan}
\date{\today}
\maketitle

\begin{abstract}
We discuss the validity of generalized Debye-H\"uckel (GDH) equation proposed by Fisher {\itshape et al.} from the functional integral point of view. The GDH theory considers fluctuations around prescribed densities of positive and negative charges. Hence we first formulate a density functional integral expression for the canonical system of Coulomb gas, and also demonstrate that this is a dual form to the Sine-Gordon theory. Our formalism reveals the following: (i) The induced charge distribution around supposed density favors not only the cancellation of additional electrostatic potential like the original DH theory, but also the countervailing of chemical potential difference between imposed and equilibrium value. (ii) As a consequence apparent charge, absent in the GDH equation, comes out in our generalized equation. (iii) That is, the GDH equation holds only in special cases.
\end{abstract}


Much attention in recent years has been paid to novel phenomena of ionic fluids and charged colloids, such as unexpected behaviors of critical exponents in ionic fluids \cite{criticality}, (micro)phase separation of charged colloids \cite{separation}, attraction between similarly charged objects \cite{attraction}, and so on. Of the numerous theoretical attempts for explaining these, this paper will be particularly concerned with a relevant tool for studying the ionic criticality: generalized Debye-H\"uckel (GDH) theory proposed by Fisher {\itshape et al.} \cite{GDH1,GDH2}. Then let us first see what is {\itshape generalized}. 

One will find two extensions from the GDH equations (\ref{GDH1}) and (\ref{GDH2}) given below. Before doing so, however, the setting is to be described: Consider here the restricted primitive model (RPM)\cite{comment}, consisting of two oppositely charged, but otherwise identical, sets of $N\equiv N_+=N_-$ hard spheres of diameter $a$ and charge per particle $\pm e$, immersed in a medium of dielectric constant $\epsilon$. Also denote by $\Psi_+({\bf r};{\bf r}_1)$ the mean electrostatic potential in the $k_BT/e$ unit (the others being the same) at a general point ${\bf r}$ when the positive charge 1 is fixed at ${\bf r}_1$, and by $\Psi_{imp}({\bf r})$ the imposed electrostatic potential determined from prescribed densities of positive and negative charges, i.e., $n_+({\bf r})$ and $n_-({\bf r})$ as $\nabla^2\Psi_{imp}({\bf r})=-4\pi l_B\,Q({\bf r})$ with putting that $l_B\equiv e^2/\epsilon k_BT$ and $Q=n_+-n_-$.

Fisher {\itshape et al.} advocate \cite{GDH1,GDH2} that the "local induced potential" defined by $\widetilde{\Psi}_+({\bf r};{\bf r}_1)\equiv\Psi_+({\bf r};{\bf r}_1)-\Psi_{imp}({\bf r})$ be to satisfy the GDH equations as follows:
  \begin{eqnarray}
  \label{GDH1}
  \nabla^2\widetilde{\Psi}_+({\bf r};{\bf r}_1)&=&-4\pi l_B[\delta({\bf r}-{\bf r}_1)-Q({\bf r})], \quad|{\bf r}-{\bf r}_1|\leq a,\\
  \label{GDH2}
  \nabla^2\widetilde{\Psi}_+({\bf r};{\bf r}_1)&=&\kappa^2({\bf r})\widetilde{\Psi}_+({\bf r};{\bf r}_1),\qquad\qquad\quad\;\;\;|{\bf r}-{\bf r}_1|\geq a,
  \end{eqnarray}
implying modifications that are in order:
\begin{itemize}
\item A remarkable difference from the original DH equation \cite{ODH} is seen in Eq. (\ref{GDH2}) where the Debye-H\"uckel screening length $\kappa^{-1}$ is generalized to be spatially dependent on the supposed densities as $\kappa^2\equiv 4\pi l_B[\, n_+({\bf r}) +n_-({\bf r}) \,]$.
\item There also appears, unlike the primary version, the second term on the right hand side of (\ref{GDH1}) representing an effective "cavity source" term.
\end{itemize}
The validity of these two generalizations are what we would like to investigate using the language of density functional integral.

For convenience of later discussion, let us further detail the GDH theory \cite{GDH1,GDH2} exploiting the above equations (\ref{GDH1}) and (\ref{GDH2}). According to this theory, the Helmholtz free energy $F$ (in the $k_BT$ unit) as sum of the imposed and induced free energy, $F=F_{imp}+F_{ind}$, is given in the following density functional form:
  \begin{eqnarray}
  \label{imp}
  F_{imp}&=&\int d{\bf r} d{\bf r'} \;\frac{l_B}{2}\,Q({\bf r})\frac{1}{|{\bf r}-{\bf r}'|}Q({\bf r'})
  +\int d{\bf r} \; n_+\ln n_++ n_-\ln n_-- n_+- n_-,
  \\
  \label{ind}
  F_{ind}&=&\sum_{\sigma=+,-}\int d{\bf r_1}\;n_{\sigma}({\bf r}_1)\int^1_0{d\lambda}\;\Phi_{\sigma}({\bf r}_1;\lambda e),
  \end{eqnarray}
where $\Phi_{\sigma}({\bf r}_1;\lambda e)$ is the mean {\itshape induced} electrostatic potential at the site ${\bf r}_1$ of a fixed ion defined as $\Phi_{\sigma}({\bf r}_1; e)\equiv\lim_{{\bf r}\to{\bf r_1}}\left[\widetilde{\Psi}_{\sigma}({\bf r}; {\bf r_1})-\sigma{l_B}/{|{\bf r}-{\bf r_1}|}\right]\;(\sigma=+,-)$. The GDH theory especially imposes $ n_{\sigma}$ $(\sigma=+,-)$ on such a simple undulation as
  \begin{equation}
  \label{imposed}
   n_{\sigma}=\bar{ n}\,[1+\Lambda\,\cos ({\bf k}\cdot{\bf r}+\theta_{\sigma})].
  \end{equation}
Consequently the quadratic terms of $\Lambda$ in the free energy difference, $F(\{ n_{\sigma}\})-F(\{\bar{ n}\})$, gives the Fourier transform of various correlation functions by taking either phase, $\Delta\theta\equiv |\theta_+-\theta_-|=0\;\mbox{or}\;\pi$, on a case-by-case basis. Thus this approach, in contrast to the original DH analysis, has succeeded in yielding both charge-charge oscillatory correlations at high densities (where $\Delta\theta=\pi$ is taken) and density-density correlations (where $\Delta\theta=0$) that exhibit a divergent correlation length at criticality.

Indeed, the validity of the GDH theory is strongly suggested not only by the above usefulness of this method, but also from satisfying both the Stillinger-Lovett sum rule \cite{GDH1,GDH2} and the exact low-density limiting-law for various correlation lengths \cite{correlation}. In terms of functional integral, however, the GDH equations (\ref{GDH1}) and (\ref{GDH2}) are {\itshape never} trivial, due to the irrelevance of the Sine-Gordon (SG) mapping \cite{Edwards}. The details follow: With use of the chemical potentials $\mu_{\sigma}$ $(\sigma=+,-)$ and the charge density operator $\hat{q}({\bf r})$ defined as $\hat{q}\equiv\hat{\rho}_+-\hat{\rho}_-$ by the number density operators $\hat{\rho}_{\pm}\equiv\sum_{i=1}^{N} \delta[{\bf r}-{\bf r}_i^{\pm}]$, the grand partition function $\Xi$ reads
  \begin{equation}
  \label{grand}
  \Xi=\sum_{N}^{\infty}\frac{e^{\mu_+ N}}{N!}\sum_{N}^{\infty}
  \frac{e^{\mu_- N}}{N!}\,Z,
  \end{equation}
where
  \begin{equation}
  \label{cano}
  Z\equiv\int d\,\{{\bf r}^{2N}\}\;\exp\left[\,-\frac{l_B}{2}\int d{\bf r} d{\bf r'}
  \;\frac{\hat{q}({\bf r})\hat{q}({\bf r'})}{|{\bf r}-{\bf r}'|}\right]
  \end{equation}
with putting that $d\{{\bf r}^{2N}\}\equiv\prod_{i=1}^{N}d{\bf r}_i^+\prod_{j=1}^{N}d{\bf r}_i^-$. In the conventional SG mapping, eqs. (\ref{grand}) and (\ref{cano}) are Hubbard-Stratonovich transformed \cite{Justin} to the functional integral of {\itshape potential field}, not of density. Therefore, within the SG frame, it is impossible to evaluate fluctuations around imposed number densities (not around prescribed potential) as the GDH theory does.

To generalize the original Debye-H\"uckel equation as Fisher {\itshape et al.} propose, it is hence indispensable to transform expressions (\ref{grand}) and (\ref{cano}) to the density functional integral formalism of the RPM. The former part in the remainder is then devoted to both its formulation and comparison with the SG theory. In the final, this formalism will reveal that the GDH equations (\ref{GDH1}) and (\ref{GDH2}) hold only in special cases.

Recently we have shown \cite{PREfru} that a dual method to the above Hubbard-Stratonovich way enables one to transform the configurational integral expressions (\ref{grand}) and (\ref{cano}) to the following density functional integral representation:
  \begin{eqnarray}
  \label{grand3}
  &&\Xi=\prod_{\sigma=+,-}\int{D}\rho_{\sigma}
  \exp\;(-H_{\Xi}\,\{\rho_{\pm}\}\,),\\
  \label{hamil3}
  &&H_{\Xi}\,\{\rho_{\pm}\}
  \equiv\frac{l_B}{2}\int d{\bf r} d{\bf r'}
  \;\frac{q({\bf r})q({\bf r'})}{|{\bf r}-{\bf r}'|}
  +\sum_{\sigma=+,-}\int d{\bf r}\;\rho_{\sigma}\ln\rho_{\sigma}
  -\rho_{\sigma}-\rho_{\sigma}\mu_{\sigma},
  \end{eqnarray}
where $D\rho_{\sigma}\propto\prod_{\{\bf r\}}d\rho_{\sigma}$, and $q$ and $\rho_{\sigma}$ are c-numbers of the corresponding operators.

What is required in eqs. (\ref{imp}) and (\ref{ind}), however, is the Helmholtz free energy. We then move to the canonical partition function $Z$, via performing the following contour integral,
  \begin{equation}
  \label{gtoc}
  Z=\prod_{\sigma=+,-}\frac{1}{2\pi i}
  \oint d\lambda_{\sigma} \frac{\Xi}{\lambda_{\sigma}^{N+1}},
  \end{equation}
where $\lambda_{\sigma}=e^{\mu_{\sigma}}$ is now a complex variable. Substituting eqs. (\ref{grand3}) and (\ref{hamil3}) into the integrand, we have
  \begin{eqnarray}
  \label{canfunc}
  Z&=&\prod_{\sigma=+,-}\int{D}\rho_{\sigma}\exp\left[-H_{\Xi}\,\{\rho_{\pm}\}-\sum_{\sigma=+,-}\int d{\bf r}\mu_{\sigma}\rho_{\sigma}\right]
\frac{1}{2\pi i}\oint d\lambda_{\sigma}\;
  \lambda_{\sigma}^{-1-[N-\int{dx}\rho_{\sigma}(x)]}.
  \end{eqnarray}
Since one finds from the Cauchy's integral theorem
  \begin{eqnarray}
  \frac{1}{2\pi i}\oint d\lambda_{\sigma}\;
  \lambda_{\sigma}^{-1-[N-\int{dx}\rho_{\sigma}(x)]}\nonumber\\
  \qquad=\left\{
  \begin{array}{ll}
  1 & \qquad\mbox{if  $\;\int{dx}\,\rho_{\sigma}(x)=N$}\\
  0 & \qquad\mbox{otherwise},
  \end{array}
  \right.
  \end{eqnarray}
eq. (\ref{canfunc}) is formally rewritten as
  \begin{eqnarray}
  \label{can2}
  Z&=&\prod_{\sigma=+,-}\int{D}\rho_{\sigma}\exp\left[\,-H_{Z}\{\rho_{\pm}\}\;\right]
\delta\left[\int d{\bf r}\rho_{\sigma}({\bf r})-N\;\right],\\
  \label{hamil2}
  H_{Z}&\equiv&\frac{l_B}{2}\int d{\bf r} d{\bf r'}
  \;\frac{q({\bf r})q({\bf r'})}{|{\bf r}-{\bf r}'|}
  +\sum_{\sigma=+,-}\int d{\bf r}\;\rho_{\sigma}\ln\rho_{\sigma}
  -\rho_{\sigma}.
  \end{eqnarray}
We have thus established the framework of which {\itshape a priori} use \cite{Caprio,Zhou,Frusawa} has been made so far.
  
Before entering into the main question, let us also see that the density functional integral formalism gives back the SG theory of the canonical version. To this end we first introduce electrostatic potential field $\psi({\bf r})$ via using the equivalence between eqs. (\ref{can2}) and (\ref{hamil2}) and the following form:
  \begin{eqnarray}
  \label{canSG}
  &&Z=\int{D}\psi\prod_{\sigma=+,-}\int{D}\rho_{\sigma}
  \exp\left[\,-H_{SG}\{\rho_{\pm};\psi\}\;\right]
  \;\delta\left[\int d{\bf r}\rho_{\sigma}({\bf r})
  -N\;\right],
  \\
  \label{hamilSG}
  &&H_{SG}\{\rho_{\pm};\psi\}\equiv\int d{\bf r}\;
  -\frac{1}{8\pi l_B}(\nabla\psi)^2
  +(\rho_+-\rho_-)\psi
  +\sum_{\sigma=+,-}\int d{\bf r}\;\rho_{\sigma}\ln\rho_{\sigma}
  -\rho_{\sigma}.
  \end{eqnarray}
In these expressions, quadratic fluctuations of the density field around the saddle-point path $\{\rho_{\pm}^{sp}\}$ are negligible as has been shown elsewhere \cite{PREfru}. Therefore Gaussian approximation to $\{\rho_{\pm}\}$ reduces to the substitution of the Boltzmann distribution, $\rho_{\sigma}^{sp}=\rho_{\sigma}^0\exp(-\sigma\psi)\,(\sigma=+,-)$, into eq. (\ref{hamilSG}), yielding
  \begin{eqnarray}
  &&Z=\int{D}\psi\,\exp\left[\,-H_{SG}\{\rho^{sp}_{\pm};\psi\}\;\right],\\
  &&H_{SG}\{\rho^{sp}_{\pm};\psi\}=-\int d{\bf r}
  \frac{1}{8\pi l_B}(\nabla\psi)^2
  +\sum_{\sigma=+,-}\rho_{\sigma}^0 \;e^{-\sigma\psi},
  \end{eqnarray}
where the total number invariance leads to $\rho_{\sigma}^0=N/\int d{\bf r}e^{-\sigma\psi}$. The above transformation demonstrates that our formalism, eqs. (\ref{can2}) and (\ref{hamil2}), is a dual expression to the canonical SG theory, though there also exists a merit that our Hamiltonian (\ref{hamil2}) has two arguments (number densities of positive and negative charges) unlike the SG theory.

We now return to the starting formulae, (\ref{can2}) and (\ref{hamil2}), and consider the effect of quadratic fluctuations around the prescribed number densities $n_{\pm}$ on the electrostatic potential. Setting that $\widetilde{\rho}_{\sigma}=\rho_{\sigma}-n_{\sigma}\,(\sigma=+,-)$ and $\widetilde{q}=\widetilde{\rho}_+-\widetilde{\rho}_-$, and separating the imposed free energy $F_{imp}$, we have
  \begin{eqnarray}
  \label{rescan}
  &&Z=e^{-F_{imp}}\prod_{\sigma=+,-}\int{D'}\widetilde{\rho}_{\sigma}\;
  e^{-\widetilde{H}_{Z}\{\widetilde{\rho}_{\pm}\}}
  \,\delta\left[\int d{\bf r}\,\widetilde{\rho}_{\sigma}({\bf r})\;\right],
  \\
  \label{reshamil}
  &&\widetilde{H}_{Z}\{\widetilde{\rho}_{\pm}\}\equiv
  \frac{l_B}{2}\int d{\bf r} d{\bf r'}
  \;\frac{\widetilde{q}({\bf r})\widetilde{q}({\bf r'})}
  {|{\bf r}-{\bf r}'|}
  +\sum_{\sigma=+,-}\int d{\bf r}\;
  \frac{\widetilde{\rho}_{\sigma}^2}{2n_{\sigma}}+
  \widetilde{\rho}_{\sigma}(\sigma\Psi_{imp}+\ln n_{\sigma}),
  \end{eqnarray}
where fixing positive charge 1 at ${\bf r}_1$ is taken into account by changing the definition of the integral measure as $D'\widetilde{\rho}_{\sigma}\propto\prod_{|{\bf r}_1-{\bf r}|\geq a}d\widetilde{\rho}_{\sigma}$, following the treatment of Fisher {\itshape et al.}\cite{GDH1,GDH2}. With use of expressions (\ref{rescan}) and (\ref{reshamil}), however, Gaussian integration over $\widetilde{\rho}_{\sigma}$ is not straightforward due to the spatial dependence of the prescribed densities $n_{\sigma}(\sigma=+,-)$ other than conventional cases \cite{Caprio}.

We then take a detour to introduce the induced potential $\widetilde{\psi}\equiv\psi-\Psi_{imp}$ similarly to eqs. (\ref{canSG}) and (\ref{hamilSG}):
  \begin{eqnarray}
  \label{rescanSG}
  &&Z=e^{-F_{imp}}\int{D}\widetilde{\psi}
  \prod_{\sigma=+,-}\int d\nu_{\sigma}\int{D}'\widetilde{\rho}_{\sigma}\;
  e^{-\widetilde{H}_{SG}\{\tilde{\rho}_{\pm};\nu_{\pm};\tilde{\psi}\}},
  \nonumber\\
  \\
  \label{reshamilSG}
  &&\widetilde{H}_{SG}\{\widetilde{\rho}_{\pm};\nu_{\pm};\widetilde{\psi}\}
  \equiv
  \int d{\bf r}\;-\frac{1}{8\pi l_B}(\nabla\widetilde{\psi})^2
  +(\widetilde{\rho}_+-\widetilde{\rho}_-)\,\widetilde{\psi}\nonumber\\
  &&\quad+\sum_{\sigma=+,-}\int' d{\bf r}\;
  \frac{\widetilde{\rho}_{\sigma}^2}{2n_{\sigma}}+
  \widetilde{\rho}_{\sigma}(\sigma\Psi_{imp}+\ln n_{\sigma})
  -\int d{\bf r}\;i\widetilde{\rho}_{\sigma}\nu_{\sigma},
  \end{eqnarray}
where use has been made of the identity that $\delta\,[\,\int d{\bf r}\;\widetilde{\rho}_{\sigma}({\bf r})\,]=\int d\nu_{\sigma}\exp(i\nu_{\sigma}\int d{\bf r}\widetilde{\rho}_{\sigma})$, and $\int'd{\bf r}$ denotes the integration with omitting the region $\Delta r_1\equiv|{\bf r}_1-{\bf r}|\leq a$. In these expressions Gaussian integration over $\widetilde{\rho}_{\sigma}$ and $\nu_{\sigma}$ becomes trivial, though the resulting form is somewhat complicated:
  \begin{eqnarray}
  \label{rescanSG2}
  &&Z=e^{-F_{imp}}\int{D}\widetilde{\psi}\;
  e^{-\widetilde{H}_{SG}\{\tilde{\rho}_{\pm}^{sp};\nu^{sp};\tilde{\psi}\}},
  \\
  \label{reshamilSG2}
  &&\widetilde{H}_{SG}\{\widetilde{\rho}_{\pm}^{sp};\nu_{\pm}^{sp};\widetilde{\psi}\}\nonumber\\
  &&=\int d{\bf r}\;-\frac{(\nabla\widetilde{\psi})^2}{8\pi l_B}
  +[\,\delta({\bf r}-{\bf r_1})-Q({\bf r})\,]\;\gamma(\Delta r_1)\,\widetilde{\psi}\nonumber\\
  &&\qquad-\sum_{\sigma=+,-}\int' d{\bf r}\;
  \frac{n_{\sigma}}{2}(\sigma\widetilde{\psi}
  +\sigma\Psi_{imp}+\ln n_{\sigma})^2\nonumber\\
  &&+\sum_{\sigma=+,-}\frac{1}{2}
  \frac{\left\{\,-\widetilde{\rho}_{\sigma}^{fix}+\int' d{\bf r}\;n_{\sigma}(\sigma\widetilde{\psi}
  +\sigma\Psi_{imp}+\ln n_{\sigma})\right\}^2}{\int' d{\bf r}\;n_{\sigma}},
  \nonumber\\
  \end{eqnarray}
where the step function $\gamma(\Delta r_1)$ is 1 for $\Delta r_1\leq a$ and $0$ otherwise, and we define $\widetilde{\rho}_{\sigma}^{fix}\;(\sigma=+,-)$ as $\widetilde{\rho}_+^{fix}\equiv[\,\delta({\bf r}-{\bf r_1})-n_+({\bf r})\,]\gamma(\Delta r_1)$ and $\widetilde{\rho}_-^{fix}\equiv-n_-({\bf r})\,\gamma(\Delta r_1)$.

Equations (\ref{rescanSG2}) and (\ref{reshamilSG2}) imply that the saddle-point paths satisfy the following relations:
  \begin{equation}
  \label{resdensity}
  \widetilde{q}^{sp}
  =-\sum_{\sigma=+,-}\sigma n_{\sigma}(\sigma\widetilde{\psi}
  +\sigma\Psi_{imp}+\ln n_{\sigma}-i\nu_{\sigma}^{sp}),
  \end{equation}
and
  \begin{equation}
  \label{chem}
  i\nu_{\sigma}^{sp}=\frac{-\widetilde{\rho}_{\sigma}^{fix}
  +\int' d{\bf r}\;n_{\sigma}(\sigma\widetilde{\psi}
  +\sigma\Psi_{imp}+\ln n_{\sigma})}{\int' d{\bf r}\;n_{\sigma}},
  \end{equation}
where $\widetilde{q}^{sp}=\widetilde{\rho}_+^{sp}-\widetilde{\rho}_-^{sp}$. We can easily check that these reduce to $\widetilde{q}^{sp}=-2\bar{n}\widetilde{\psi}$ and $i\nu_{\sigma}^{sp}=\ln\bar{n}$ in the conventional case of $n_{\sigma}=\bar{n}$. Such reduction demonstrates that our formulation correctly includes the original DH theory \cite{ODH}, and that $i\nu_{\sigma}^{sp}\, (\sigma=+,-)$ merely correspond to the chemical potentials $\mu_{\sigma}$ in equilibrium.

Thus expression (\ref{resdensity}) with (\ref{chem}) provides the physical insight into fluctuations around prescribed densities as follows: in general the induced charge density distribution $\widetilde{q}^{sp}$ around supposed density favors not only the cancellation of additional electrostatic potential $\widetilde{\psi}$ like the original DH theory, but also the countervailing of the chemical potential difference between imposed and equilibrium value,
   \begin{equation}
   \label{adchem}
   \Delta\mu_{\sigma}\equiv\sigma\Psi_{imp}+\ln n_{\sigma}-i\nu_{\sigma}^{sp}.
   \end{equation}

Finally let us write down the DH-like equations generalized via density functional integral formalism. Saddle-point approximation to eqs. (\ref{rescanSG2}) and (\ref{reshamilSG2}) yields
  \begin{eqnarray}
  \label{ourGDH1}
  &&\nabla^2\widetilde{\Psi}_+({\bf r};{\bf r}_1)=-4\pi l_B[\delta({\bf r}-{\bf r}_1)-Q({\bf r})],\qquad\qquad|{\bf r_1}-{\bf r}|\leq a,
  \\
  \label{ourGDH2}
  &&\nabla^2\widetilde{\Psi}_+({\bf r};{\bf r}_1)
  =\kappa^2({\bf r})\widetilde{\Psi}_+({\bf r};{\bf r}_1)
  -4\pi l_B\,q_{ap}({\bf r}),\qquad\;
  |{\bf r_1}-{\bf r}|\geq a,
  \end{eqnarray}
where the saddle-point path of $\{\widetilde{\psi}\}$ is identified with $\widetilde{\Psi}_+$ used in the GDH equations (1) and (2), and $q_{ap}\equiv-\sum_{\sigma=+,-}\sigma n_{\sigma}\Delta\mu_{\sigma}$ is the apparent charge arising from the chemical potential difference $\Delta\mu_{\sigma}$ given by (\ref{adchem}). Note also that eq. (\ref{ourGDH2}) corresponds to the equation that $\nabla^2\widetilde{\Psi}_+=-4\pi l_B\widetilde{q}^{sp}$ with use of eq. (\ref{resdensity}). Comparing our equations (\ref{ourGDH1}) and (\ref{ourGDH2}) with the GDH ones (\ref{GDH1}) and (\ref{GDH2}), one immediately finds that the former equations are different from the latter, in that the apparent charge $q_{ap}$ comes out on the right hand side of (\ref{ourGDH2}), though both the spatially dependent screening length $\kappa^{-1}(\bf r)$ and the cavity source term [$4\pi l_B Q$ in eq. (\ref{ourGDH1})] can be reproduced indeed.

Accordingly the remaining problem is to investigate when the apparent charge $q_{ap}$ in eq. (\ref{ourGDH2}) disappears. A trivial condition is the chemical equilibrium $\Delta\mu_{\sigma}=0$ where the imposed densities satisfy the Boltzmann distribution $n_{\sigma}=\exp(\mu_{\sigma}-\sigma\Psi_{imp})$. In such cases, though, what is imposed is not densities but potential, and the associated equations, already verified \cite{Zhou,Frusawa,Fixman,MPB}, are directly obtainable from the conventional SG theory, not via our formalism; this is not the case with us. A relevant case is to impose $n_{\sigma}$ on relation (\ref{imposed}) with $\Delta\theta=0$ as the GDH theory does, where $\Psi_{imp}=0$ and the apparent charge $q_{ap}$ is negligible under the constraint that $\nu_+^{sp}=\nu_-^{sp}$.

We thus conclude from the functional integral point of view the following: The GDH equations (1) and (2) proposed by Fisher {\itshape et al.} are valid for the prescribed densities of $n_{\sigma}=\bar{n}[1+\Lambda\cos({\bf k}\cdot{\bf r})]$, i.e., $\Delta\theta=0$, where the density-density correlation length has been extracted \cite{GDH1}. However, in the other case of $n_{\sigma}=\bar{n}[1+\sigma\Lambda\cos({\bf k}\cdot{\bf r})]\,(\sigma=+,-)$, i.e., $\Delta\theta=\pi$, where both the charge-charge correlation length and the Lebowitz length have been derived \cite{GDH1,correlation}, we have
  \begin{equation}
  q_{ap}\approx
  2\bar{n}\Psi_{imp}+Q+{\mathcal O}[\Lambda^2],
  \end{equation}
and hence apparent charge is to be considered explicitly in the calculation of $\widetilde{\Psi}_+$; evaluating to what extent the additional term changes the previous results \cite{GDH1,correlation} remains a future problem.

\bigskip
We acknowledge the financial support from the Ministry of Education, Science, Culture, and Sports of Japan.



\begin{references}
\bibitem{comment}
In ref. \cite{GDH2}, one component plasma (OCP) is considered; however, extending the present discussion from the RPM to the OCP is straightforward.
\end{references}
\end{document}